\nonstopmode
\documentclass[copyright,creativecommons]{eptcs} 
\providecommand{\event}{FICS 2010}
\usepackage{breakurl} 
\usepackage{ifthen}
\usepackage{amsmath}
\usepackage{amssymb}
\usepackage{stmaryrd}
\usepackage{xspace}

\usepackage[usenames,dvipsnames]{color}
\usepackage{hyperref}
\hypersetup{colorlinks=true, linkcolor=Red, citecolor=ForestGreen, urlcolor=RoyalBlue}

\usepackage{listings}
\usepackage{alltt}
 
\newcommand{\defHaskelllistings}{%
  \lstset{%
    language=Haskell,%
    basicstyle=\ttfamily\small,
    keywordstyle=\ttfamily\bfseries,
    identifierstyle=,%
    commentstyle=\itshape,%
    columns=flexible,
    showstringspaces=false,%
    xleftmargin=\codeindent,
    breaklines=true,%
    deletekeywords={succ,zero,head,tail,zipWith,Either,List},%
    morekeywords={Set,Size,fun,cofun,pattern},
    literate={\\}{{$\lambda$}}1 {->}{{$\rightarrow$~}}2
             {<=}{{$\leq$~}}2 {<}{{$<$~}}1
     }%
}
\defHaskelllistings

\ifdefined\LONGVERSION
  \relax
\else
\newcommand{\LONGVERSION}[1]{}

\fi

\newcommand{\foetus}{\ensuremath{\mathsf{foetus}}\xspace}

\newcommand{\Fomega}{\ensuremath{\mathsf{F}^\omega}}
\newcommand{\der}{\,\vdash}

\def\lv{\mathopen{{[\kern-0.14em[}}}    
\def\rv{\mathclose{{]\kern-0.14em]}}}   

\newcommand{\dens}[1]{\mathopen{[\kern-0.3ex[}#1\mathclose{]\kern-0.3ex]}}
\newcommand{\denk}[2]{\mathopen{\{\kern-0.3ex|}#1\mathclose{|\kern-0.3ex\}}_{#2}}
\def\lo{\mathopen{{\lceil\kern-0.25em\lceil}}}    
\def\ro{\mathclose{{\rfloor\kern-0.25em\rfloor}}}

\def\ltox#1{\buildrel\raise1pt\hbox{$\scriptstyle#1$}\over\longrightarrow}

\def\tocolow{\buildrel\raise-5pt\hbox{$\scriptscriptstyle+$}\over\rightarrow}

\newcommand{\abbrev}[1]{#1} 
\newcommand{\cf}{\abbrev{cf.}\@\xspace} 
\newcommand{\eg}{\abbrev{e.\,g.}}

\newcommand{\ie}{\abbrev{i.\,e.}}

\newcommand{\etal}{\abbrev{et.\,al.}\@\xspace}

\newcommand{\para}[1]{\paragraph*{\it#1}}
\newcommand{\paradot}[1]{\para{#1.}}

\newcommand{\mgoal}[1][]{\mbox{goal\ifthenelse{\equal{#1}{}}{}{~#1}}}

\newcommand{\ru}[2]{\dfrac{\begin{array}[b]{@{}c@{}} #1 \end{array}}{#2}}

\newcommand{\tfix}{\mathsf{fix}}

\newcommand{\tlet}{\mathsf{let}}
\newcommand{\tin}{\mathsf{in}}
\newcommand{\letin}[2]{\tlet~#1=#2~\tin~}

\newcommand{\mA}{{{}'\kern-.6ex A}}
\newcommand{\mB}{{{}'\kern-.3ex B}}
\newcommand{\prequote}[1]{{{}'\kern-.3ex #1}}

\newcommand{\FORALL}{\bm\forall\kern-1.38ex\bm\forall}
\newcommand{\TIMES}{\mathop{\bm{\times}\kern-1.93ex\bm{\times}}}
\newcommand{\EXISTS}{\bm\exists\kern-1.35ex\raisebox{0.03em}{\ensuremath{\bm\exists}}}
\newcommand{\ordto}{\mathord{\to}}
\newcommand{\TO}{\mathop{\bm\ordto\kern-2.55ex\raisebox{0.03em}{\ensuremath{\bm\ordto}}}}

\newcommand{\Bool}{\mathsf{Bool}}
\newcommand{\ttrue}{\mathsf{true}}

\newcommand{\tid}{\mathsf{id}}

\newcommand{\EL}{\mathcal{E}\kern-0.2ex\ell}

\newcommand{\List}{\mathsf{List}}
\newcommand{\CoList}{\mathsf{CoList}}
\newcommand{\tcons}{\mathsf{cons}}

\newcommand{\Size}{\mathsf{Size}}
\newcommand{\Stream}{\mathsf{Stream}}

\newcommand{\vsa}{\mathit{sa}}
\newcommand{\vsb}{\mathit{sb}}
\newcommand{\vsc}{\mathit{sc}}
\newcommand{\tzipWith}{\mathsf{zipWith}}
\newcommand{\vas}{\mathit{as}}
\newcommand{\vbs}{\mathit{bs}}
\newcommand{\SP}{\mathsf{SP}}
\newcommand{\tget}{\mathsf{get}}
\newcommand{\tput}{\mathsf{put}}
\newcommand{\SPAB}{\mathsf{SP}~A~B}
\newcommand{\vsp}{\mathit{sp}}

\newcommand{\MU}{\overline\mu}

\newcommand{\x}{\mathsf{x}}
\newcommand{\xdel}[1][]{\ifthenelse{\equal{#1}{}}{\x_\Delta}{\x_{\Delta+#1}}}

\newcommand{\T}{\mathcal{T}}

\newcommand{\VDash}{\mathrel{\mathord{|}\kern-0.15ex\mathord{\models}}}

\def\squareforqed{\ensuremath{\Box}}
\def\qed{\ifmmode\squareforqed\else{\unskip\nobreak\hfil
\penalty50\hskip1em\null\nobreak\hfil\squareforqed
\parfillskip=0pt\finalhyphendemerits=0\endgraf}\fi}

\newenvironment{proof*}[1][]{\noindent\ifthenelse{\equal{#1}{}}{{\it
      Proof.}}{{\it Proof #1.}}\hspace{2ex}}{\bigskip}

\newlength{\codeindent}
\newenvironment{code}{\newline\noindent\begin{minipage}{\textwidth}}{\end{minipage}}

\newenvironment{quotecode}{%
  \begin{list}{}{%
    \setlength{\topsep}{0pt}%
    \setlength{\leftmargin}{\codeindent}%
    \setlength{\itemindent}{0pt}%
  }%
  \item[]}{\end{list}}

\title{Type-Based Termination, Inflationary Fixed-Points, \\
  and Mixed Inductive-Coinductive Types}
\def\titlerunning{Type-Based Termination, Inflationary Fixed-Points,
  and Mixed Inductive-Coinductive Types}
\author{Andreas Abel
  \institute{
    Department of Computer Science \\
    Ludwig-Maximilians-University Munich, Germany
  }%
  \email{andreas.abel@ifi.lmu.de}%
}
\def\authorrunning{Andreas Abel}

\begin{document}
\maketitle

\begin{abstract}\noindent
  Type systems certify program properties in a compositional way.
  From a bigger program one can abstract out a part and certify the
  properties of the resulting abstract program by just using the type
  of the part that was abstracted away.  \emph{Termination} and
  \emph{productivity} are non-trivial yet desired program properties, and
  several type systems have been put forward that guarantee
  termination, compositionally.  These type systems are intimately
  connected to the definition of least and greatest fixed-points by
  ordinal iteration.  While most type systems use ``conventional''
  iteration, we consider inflationary iteration in this article.  We
  demonstrate how this leads to a more principled type system, with
  recursion based on well-founded induction.  The type system has a
  prototypical implementation, MiniAgda, and we show in particular 
  how it certifies
  productivity of corecursive and mixed recursive-corecursive 
  functions.
\end{abstract}

%
%
%
%
%
%

\section{Introduction: Types, Compositionality, and Termination}
\label{sec:intro}



While basic types like \emph{integer}, \emph{floating-point number},
and \emph{memory address} arise on the machine-level of most current
computers, higher types like function and tuple types are abstractions
that classify values.  Higher types serve to guarantee certain good
program behaviors, like the classic ``don't go wrong'' absence of
runtime errors \cite{milner:goWrong}.  Such properties are usually not
compositional, \ie, while a function $f$ and its argument $a$ might
both be well-behaved on their own, their application $f\,a$ might
still go wrong.  This issue also pops up in termination proofs: take
$f = a = \lambda x.\, x\,x$, then both are terminating, but their
application loops.  To be compositional, 
the property \emph{terminating} needs to be
strengthened to what is often called \emph{reducible} 
\cite{girard:thesis}
or \emph{strongly computable} \cite{tait:functionalsFiniteTypeI},
leading to a semantic notion of type.  While the bare properties
are not compositional, \emph{typing} is.

Type \emph{polymorphism}
\cite{reynolds:systemF,girard:thesis,milner:goWrong}
has been invented for compositionality in the opposite direction: We
want to decompose a larger program into smaller parts such that the
well-typedness of the parts imply the well-typedness of the whole
program.  Consider  $(\lambda x.x) \, (\lambda x.x)\, \ttrue$,
a simply-typed program
which can be abstracted to $\letin \tid {\lambda x.x} {\tid~\tid~\ttrue}$.
The two occurrences of $\tid$ have different type, namely $\Bool \to
\Bool$ and $(\Bool \to \Bool) \to \Bool \to \Bool$, and the easiest
way to type check the new program is to just inline the definition of
$\tid$.  This trick does not scale, however, making type checking infeasible
and separate compilation of modules impossible.  The accepted solution is to
give $\tid$ the polymorphic type $\forall X.\, X \to X$ which can be
instantiated to the two required types of $\tid$.

Termination checking, if it is to scale to software development with
powerful abstractions, needs to be compositional. Just
like for other non-standard analyses, \eg, strictness, resource
consumption and security, type-based termination promises to be a
model of success.  Current termination checkers, however, like
\foetus~\cite{abelAlti:predicative,wahlstedt:master,altenkirchDanielsson:par10},
the one of Agda \cite{norell:PhD}, and
Coq's guardedness check~\cite{gimenez:guardeddefinitions,barras:jfla10}
are not type-based, but syntactic.  Let us see how this affects
compositionality.
Consider the following recursive program defined by pattern matching.
We use the syntax of MiniAgda \cite{abel:par10}, in this and all
following examples.
\begin{code}
\begin{lstlisting}
fun everyOther : [A : Set] -> List A -> List A
{ everyOther A nil                          = nil
; everyOther A (cons a nil)                 = nil
; everyOther A (cons a (cons a' as))        = cons a (everyOther A as)
}
\end{lstlisting}
\end{code}
The polymorphic function \lstinline|everyOther| returns a list
consisting of every second element of the input list. 
Since the only recursive call happens on sublist
\lstinline|as| of the input list \lstinline|cons a (cons a' as)|,
termination is evident.  We
say that the call argument decreases in the \emph{structural order};
this order, plus lexicographic extensions, is in essence
the termination order accepted by the proof assistants Agda, Coq, and
Twelf \cite{pientka:termination}.

The function distinguishes on the empty list, the singleton list, and
lists with at least 2 elements.  Such a case distinction is used in
list sorting algorithms, too, so we may want to abstract it from
\lstinline|everyOther|. 
\begin{code}
\begin{lstlisting}
fun zeroOneMany : [A : Set] -> List A -> [C : Set] -> 
  (zero  : C) ->                     
  (one   : A -> C) ->                
  (many  : A -> A -> List A -> C) -> 
  C
{ zeroOneMany A nil                          C zero one many = zero
; zeroOneMany A (cons a nil)                 C zero one many = one a
; zeroOneMany A (cons a (cons a' as))        C zero one many = many a a' as
}
\end{lstlisting}
\end{code}
After abstracting away the case distinction, termination is no longer
evident; the program is rejected by Agda's termination checker \foetus.
\begin{code}
\begin{lstlisting}
fun everyOther : [A : Set] -> List A -> List A
{ everyOther A l = zeroOneMany A l (List A) 
    nil
    (\ a         -> nil)
    (\ a a' as   -> cons a (everyOther A as)) 
}
\end{lstlisting}
\end{code}
Whether the recursive call argument \lstinline|as| is structurally
smaller than the input \lstinline|l| depends on the definition of
\lstinline|zeroOneMany|.  In such situations, Coq's guardedness check
may inline the definition of \lstinline|zeroOneMany| and succeed.  Yet
in general, as we have discussed in the context of type checking, inlining
definitions is expensive, and in case of recursive definitions,
incomplete and brittle.  Current Coq~\cite{inria:coq83} 
may spend minutes on checking a
single definition, and fail nevertheless.

Type-based termination can handle abstraction as in the above example,
by assigning a more informative type to \lstinline|zeroOneMany| that
guarantees that the list passed to \lstinline|many| is structurally
smaller than the list analyzed by \lstinline|zeroOneMany|.  Using this
restriction, termination of \lstinline|everyOther| can be guaranteed.
To make this work, 
we introduce a purely administrative type \lstinline|Size| and let
variables \lstinline|i|, \lstinline|j|, and \lstinline|k| range over
\lstinline|Size|.  The type of lists is refined as 
\lstinline|List A i|, 
meaning lists of length $< i$.  We also add bounded size
quantification $\bigcap_{j < i} T(j)$, in concrete syntax
\lstinline|[j <  i] ->  T j|, which lets \lstinline|j| only be
instantiated to sizes strictly smaller than \lstinline|i|.  The
refined type of \lstinline|zeroOneMany| thus becomes:
\begin{code}
\begin{lstlisting}
fun zeroOneMany : [A : Set] -> [i : Size] -> List A i -> [C : Set] -> 
  (zero  : C) ->                                             
  (one   : A -> C) ->                                        
  (many  : [j < i] -> A -> A -> List A j -> C) -> 
  C
\end{lstlisting}
\end{code}
The list passed to \lstinline|many| is bounded by size \lstinline|j|, which is
strictly smaller than \lstinline|j|.  This is exactly the information
needed to make \lstinline|everyOther| termination-check.

Barthe \etal~\cite{bartheGregoirePastawski:lpar06} study type-based
termination as an automatic analysis ``behind the curtain'', with no
change to the user syntax of types.  Size quantification is restricted
to rank-1 quantifiers, known as ML-style quantification
\cite{milner:goWrong}.  This excludes the type of
\lstinline|zeroOneMany|, which has a rank-2 (bounded) quantification.
Higher-rank polymorphism is not inferable automatically, yet without
it we fall short of our aim: compositional termination.  Anyway, the
prerequisite for inference is the availability of the source code,
which fails for abstract interfaces (such as parametrized modules in
Agda, Coq, or ML).  Thus, we advocate a type system with explicit size
information based on the structural order.  
It will be presented in the remainder of this article.

\section{Sizes, Iteration, and Fixed-Points}
\label{sec:fix}

In the following, rather than syntactic we consider semantic types
such as sets of terminating terms.  
We assume that types form a complete
lattice $(\T,\subseteq,\bigcap,\bigcup)$
with least element $\bot$ and greatest element $\top$.
Further, let the usual type operators 
$+$ (disjoint sum), $\times$
(Cartesian product), and $\to$ (function type) have a sensible
definition. 
 
Inductive types $\mu F$, such as $\List~A$, are conceived as least fixed
points of monotone type constructors $F$, for lists this being $F\,X = \top
+ A \times X$. 
Constructively \cite{cousot:constructiveTarski}, least fixed
points are obtained on a $\cup$-semilattice
by ordinal iteration up to a sufficiently large
ordinal $\gamma$.  
Let $\mu^\alpha F$ denote the $\alpha$th \emph{iterate} or \emph{approximant},
which is defined by transfinite recursion on $\alpha$:
\[
\begin{array}{l@{~}lll@{\quad}l}
  \mu^0 & F & = & \bot & 
    \mbox{zero ordinal: least element of the lattice} \\
  \mu^{\alpha+1} & F & = & F\,(\mu^\alpha F) & 
    \mbox{successor ordinal: iteration step} \\
  \mu^{\lambda} & F & = & \bigcup_{\alpha<\lambda} \mu^\alpha F & 
    \mbox{limit ordinal: upper limit} 
\end{array}
\]
For monotone $F$, iteration is monotone, \ie, $\mu^\alpha F \subseteq
\mu^\beta F$ for $\alpha \leq \beta$.
At some ordinal $\gamma$, which we call \emph{closure ordinal} of
this inductive type, we have $\mu^\alpha F = \mu^\gamma F$ for all
$\alpha \geq \gamma$---the chain has become stationary, the least
fixed point has been reached.  For polynomial $F$, \ie, those
expressible without
a function space, the closure ordinal is $\omega$.  The index $\alpha$
to the approximant $\mu^\alpha F$ is a strict upper bound on the
\emph{height} of the well-founded trees inhabiting this type; in the
case of lists (which are linear trees) it is a
strict upper bound on the length.
 
Dually, coinductive types $\nu F$ are constructed on a
$\cap$-semilattice by iteration from above.
\[
\begin{array}{l@{~}lll@{\quad}l}
  \nu^0 & F & = & \top & 
    \mbox{zero ordinal: greatest element of the lattice} \\
  \nu^{\alpha+1} & F & = & F\,(\nu^\alpha F) & 
    \mbox{successor ordinal: iteration step} \\
  \nu^{\lambda} & F & = & \bigcap_{\alpha<\lambda} \nu^\alpha F & 
    \mbox{limit ordinal: lower limit} 
\end{array}
\]
Iteration from above is antitone, \ie, $\nu^\alpha F \supseteq
\nu^\beta F$ for $\alpha \leq \beta$.
The chain of approximants starts with the all-type $\top$ and descends
towards the greatest fixed-point $\nu F$.  In case of the above $F$
this would be $\CoList~A$, the type of possibly infinite lists over
element type $A$.  The index $\alpha$ in the approximant $\nu^\alpha
F$ could be called the \emph{depth} of the non-well-founded trees
inhabiting this type.  It is a lower bound on how deep we can descend
into the tree before we hit undefined behavior ($\top$).

The central idea of type-based termination, going all the way back to
Mendler~\cite{mendler:lics}, Hughes, Pareto, and
Sabry~\cite{hughesParetoSabry:popl96}, Gim\'enez~\cite{gimenez:strec},
and Amadio and Coupet-Grimal \cite{amadio:guardcondition}
is to introduce syntax to speak about approximants in the type system.
Common to the more expressible systems, such as 
Barthe \etal \cite{bartheGregoireRiba:lernet08} and
Blanqui \cite{blanqui:rta04} is syntax for ordinal variables $i$, 
ordinal successor $\mathsf{s}\,a$ (MiniAgda: \lstinline|$a|), 
closure ordinal $\infty$ (MiniAgda: \lstinline|#|) and data type
approximants $D^a$ (MiniAgda: \eg, \lstinline|List A i|).
Hughes \etal and the author \cite{abel:lmcs07} have also quantifiers
$\forall i.\, T$ over ordinals 
(MiniAgda: \lstinline|[i : Size] ->   T|).

How do we get a recursion principle from approximants?  Consider
the simplest example: constructing an infinite repetition $r$ of a
fixed element $a$ by corecursion.  After assembling the
colist-constructor $\tcons : A \to \CoList~A~i \to \CoList~A~(i+1)$ on
approximants, we give a recursive equation $r = \tcons~a~r$ with the
following typing of the r.h.s.
\[
  i : \Size,~ r : \CoList~A~i \der \tcons~a~r : \CoList~A~(i+1)
\] 
The types certify that each unfolding of the recursive definition of
$r$ increases the number of produced colist elements by one, hence, in
the limit we obtain an infinite sequence and, in particular, $r$ is productive.
Our example is a special instance of the recursion principle of
type-based termination, expressible as type assignment for the
fixpoint combinator:
\[
\ru{f ~:~ \forall i.~ T\;i \to T\;(i+1)
  }{\tfix~f ~:~ \forall i.~ T\;i}
\]
(Take $T = \CoList~A$ and $f = \lambda r.~ \tcons~a~r$ to reconstruct
the example.)  The fixed-point rule can be justified by transfinite
induction on ordinal index $i$.   While the successor case is covered
by the premise of the rule, for zero and limit case 
the size-indexed type $T$ must satisfy two conditions: 
$T~0 = \top$ (\emph{bottom check}) and  
$\bigcap_{\alpha<\lambda} T\,\alpha \subseteq T\;\lambda$ for
limit ordinals $\lambda$ \cite{hughesParetoSabry:popl96}.  
The latter condition is non-compositional,
but has a compositional generalization, \emph{upper semi-continuity}
$\bigcap_{\alpha<\lambda} \bigcup_{\alpha \leq \beta < \lambda}
T\,\beta \subseteq T\;\lambda$~\cite{abel:lmcs07}.

The soundness of type-based termination in different variants for
different type systems has been assessed in at least 5 PhD theses:
Barras~\cite{barras:PhD} (CIC),
Pareto~\cite{pareto:PhD} (lazy ML),
Frade~\cite{frade:PhD} (STL),
the author~\cite{abel:PhD} (\Fomega), and
Sacchini~\cite{sacchini:PhD} (CIC).
Recently, Barras~\cite{barras:jfr10} has completed a comprehensive
formal verification in Coq, by implementing a set-theoretical model of
the CIC with type-based termination.

However, type-based termination has not been integrated into bigger
systems like Agda and Coq.  There are a number of reasons:




\begin{enumerate}

\item \label{it:sub} Subtyping. \\
  The inclusion relation between 
  approximants gives rise to subtyping, and for dependent types,
  subtyping has not been fully explored.  While there are basic theory
  \cite{aspinallCompagnoni:tcs01,chen:subtypingCC}, substantial
  work on coercive subtyping \cite{chen:popl03,luoAdams:mscs08} 
  and new results on
  Pure Subtype Systems \cite{hutchins:popl10}, no theory of
  higher-order polarized subtyping \cite{steffen:PhD,abel:mscs06} has
  been formulated for dependent types yet.  In practice, the
  introduction of subtyping means that already complicated 
  higher-order unification has to be replaced by preunification
  \cite{qianNipkow:jar94}. 

\item \label{it:irr} Erasure.\\
  Mixing sizes into types and expressions means that one also
  needs to erase them after type checking, since they have no
  computational significance.  The type system must be able to
  distinguish relevant from irrelevant parts.  This is also work in
  progress, partial solutions have been given, \eg,  by Barras and Bernardo
  \cite{barrasBernardo:fossacs08} and the author
  \cite{abel:fossacs11}.

\item \label{it:semi} Semi-continuity.\\
  A technical condition like
  semi-continuity can kill a system as a candidate for the foundation
  of logics and programming.  It seems that it even deters the
  experts:  Most systems for type-based termination
  replace semi-continuity by a rough approximation, trading
  expressivity for simplicity---Pareto and the
  author being notable exceptions.


\item \label{it:match} Pattern matching.\\
  The literature on type-based termination is a bit thin when it comes
  to pattern matching.  Pattern matching on sized inductive types has
  only been treated by Blanqui \cite{blanqui:rta04}.  Pattern matching on
  coinductive types is known to violate subject reduction in dependent
  type theory (detailed analysis by McBride \cite{mcBride:calco09}).  
  Deep matching on sized types can lead to a surprising paradox
  \cite{abel:par10}. 

\end{enumerate}

While items~\ref{it:sub} and \ref{it:irr} require more work,
items~\ref{it:semi} and \ref{it:match} can be addressed by switching
to a different style of type-based termination, which we study in the
next section.

\section{Inflationary Iteration and Bounded Size Quantification}
\label{sec:infl}

Sprenger and Dam \cite{sprengerDam:fossacs03} note that for monotone
$F$,
\[
  \mu^\alpha F = \bigcup_{\beta<\alpha} F~(\mu^\beta F)
\]
and base their system of \emph{circular proofs in the $\mu$-calculus} on
this observation.  They introduce syntax for
unbounded $\exists i$ and bounded $\exists j<i$ ordinal existentials
and for approximants $\mu^i$ (\cf Dam and Gurov \cite{damGurov:jlc02}
and Sch\"opp and Simpson \cite{schoeppSimpson:fossacs02}).  
Induction is well-founded induction on
ordinals, and no semi-continuity is required.

A first thing to note is that if we take above equation as the
\emph{definition} for $\mu^\alpha F$, 
the chain $\alpha \mapsto \mu^\alpha F$ is monotone 
regardless of monotonicity of $F$.  This style of iteration from below
is called \emph{inflationary iteration} and the dual,
\emph{deflationary iteration},
\[
  \nu^\alpha F = \bigcap_{\beta<\alpha} F~(\nu^\beta F)
\]
always produces a descending chain.  While inflationary iteration
of $F$ becomes stationary at some closure ordinal $\gamma$, the limit
$\mu^\gamma F$ is only a pre-fixed point of $F$, \ie, $F~(\mu^\gamma
F) \subseteq \mu^\gamma F$.  This means we can construct elements in a
inflationary fixed-point as usual, but not necessarily analyze them
sensibly.  Unless $F$
is monotone, destructing an element of $\mu^\gamma F$ yields only an
element of $F\,(\mu^\beta F)$ for some $\beta < \gamma$ and not one of
$F\,(\mu^\gamma F)$.
Dually, deflationary iteration reaches a
post-fixed point $\nu^\gamma F \subseteq F~(\nu^\gamma F)$ giving the
usual destructor, but the constructor has type 
$(\forall \beta<\gamma.~F~(\nu^\beta F)) \to \nu^\gamma F$.

While we have not come across a useful application of negative
inflationary fixed points in programming, inflationary iteration leads
to ``cleaner'' type-based termination.  Inductive data constructors
have type $(\exists j<i.\; F\,(\mu^j F)) \to \mu^i F$, meaning that when
we pattern match at inductive type $\mu^i F$, we get a fresh size
variable $j < i$ and a rest of type $F\,(\mu^j F)$.  This is the
``good'' way of matching that avoids paradoxes \cite{abel:par10};
find it also in Barras \cite{barras:jfr10}.  Coinductive data has type
$\nu^i F \cong \forall j<i.\; F\,(\nu^j F)$, akin to a dependent function type.
We cannot match on it, only apply it to a size, preventing subject
reduction problems mentioned in the previous section. 
Finally, recursion becomes
well-founded recursion on ordinals,
\[
  \ru{f : \forall i.~ (\forall j<i.~T~j) \to T~i
    }{\tfix~f : \forall i.~T~i}
\]
with no condition on $T$.  Also, just like in PiSigma
\cite{alti:flops10}, 
we can dispose of inductive and coinductive types in favor of
recursion.
We just define
approximants recursively using bounded quantifiers;  for instance,
sized streams are $\Stream~A~i = \forall j<i.~ A \times \Stream~A~j$,
and in MiniAgda:
\begin{code}
\begin{lstlisting}
cofun Stream : +(A : Set) -(i : Size) -> Set 
{ Stream A i = [j < i] -> A & Stream A j
}
\end{lstlisting}
\end{code}
MiniAgda checks that $\Stream~A~i$ is monotone in element type $A$ and
antitone in depth $i$, as specified by the polarities \lstinline|+|
and \lstinline|-| in the type signature.
If we erase sizes to $()$ and $\Size$ to the non-informative type $\top$, we obtain 
$\Stream~A~() = \top \to A \times \Stream~A~()$ which is a possible
representation of streams in call-by-value languages.  Thus, size
quantification can be considered as type \emph{lifting}, size
application as \emph{forcing} and size abstraction as \emph{delaying}.
\begin{code}
\begin{lstlisting}
let tail [A : Set] [i : Size] (s : Stream A $i) : Stream A i
  = case (s i) { (a, as) -> as }   
\end{lstlisting} 
\end{code}
Taking the tail requires a stream of non-zero depth $i+1$.
Since $s ~:~ \forall j<(i+1).~ A \times \Stream~A~j$, 
we can apply it to
$i$ (\emph{force} it) and then take its second component.

Zipping two streams $\vsa = a_0,a_1,\dots$ and $\vsb = b_0,b_1,\dots$ 
with a function $f$ yields a stream $\vsc = f(a_0,b_0), f(a_1,b_1), \dots$
whose depth is the minimum of the depths of $\vsa$ and $\vsb$.  Since
depths are lower bounds, we can equally state that all three streams
have a common depth $i$.
\begin{code}
\begin{lstlisting}
cofun zipWith :    [A, B, C : Set] (f : A -> B -> C)
                   [i : Size] (sa : Stream A i) (sb : Stream B i) -> Stream C i 
{ zipWith A B C f i sa sb j = 
  case (sa j, sb j) : (A & Stream A j) & (B & Stream B j) 
  { ((a, as), (b, bs)) -> (f a b, zipWith A B C f j as bs) 
  }
}
\end{lstlisting}
\end{code}
Forcing the recursively defined stream $\tzipWith~A~B~C~f~i~\vsa~\vsb$
by applying it to $j < i$ yields a head-tail pair
$(f~a~b,~\tzipWith~A~B~C~f~j~\vas~\vbs)$ which is computed from heads
$a$ and $b$ and tails $\vas$ and $\vbs$ of the forced input streams
$\vsa~j$ and $\vsb~j$.  The recursion is well-founded since $j<i$.

The famous Haskell one-line definition
\lstinline|fib = 0 : 1 : zipWith (+) fib (tail fib)|
 of the Fibonacci
stream \lstinline|0 : 1 : 1 : 2 : 3 : 5 : 8 : 13...| can now be
replayed in MiniAgda.
\begin{code}
\begin{lstlisting}
cofun fib : [i : Size] -> |i| -> Stream Nat i
{ fib i =   \ j -> (zero, 
            \ k -> (one, 
            zipWith Nat Nat Nat add k 
              (fib k) 
              (tail Nat k (fib j))))
} 
\end{lstlisting}
\end{code} 
The \lstinline+|i|+ in the type explicitly states that ordinal $i$
shall serve as
termination measure (syntax due to Xi~\cite{xi:terminationHOSC}). Note
the two delays $\lambda j<i$ and $\lambda k<j$ and the two recursive
calls, both at smaller depth $j,k < i$.  Such a definition is beyond
the guardedness check \cite{coquand:infiniteObjects} of Agda and Coq,
but here the type system communicates that $\tzipWith$ preserves the
stream depth and, thus, productivity.

While our type system guarantees termination and productivity at
run-time, \emph{strong} normalization, in particular when reducing
under $\lambda$-abstractions, is lost when coinductive types are just
defined recursively.  Thus, equality testing of functions has to be
very intensional ($\alpha$-equality \cite{alti:flops10}), 
since testing $\eta$-equality may loop.  McBride
\cite{mcBride:calco09} suggests an extensional propositional equality
\cite{altenkirchMcBrideSwierstra:plpv07} as cure.

Having explained away inductive and coinductive types, mixing them does
not pose a problem anymore, as we will see in the next section.

\section{Mixing Induction and Coinduction}
\label{sec:mixed}

A popular mixed coinductive-inductive type are stream processors
\cite{ghaniHancockPattinson:entcs06}
given recursively by the equation
$\SP~A~B = (A \to \SP~A~B) + (B \times \SP~A~B)$.  The intention is
that  $\SP~A~B$ represents continuous functions from $\Stream~A$ to
$\Stream~B$, meaning that only finitely many $A$'s are taken from the
input stream before a $B$ is emitted on the output stream.  This
property can be ensured by nesting a least fixed-point into a greatest
one:  $\SP~A~B = \nu X.\, \mu Y.\, (A \to Y) + (B \times X)$
\cite{abel:aplas07,ghaniHancockPattinson:lmcs09}.  The greatest fixed-point
unfolds to $\mu Y.\, (A \to Y) + (B \times \SP~A~B)$, hence, whenever
we chose the second alternative, the least fixed-point is
``restarted''.  Thus, we can conceive $\SP~A~B$ by a \emph{lexicographic}
ordinal iteration
\[
  \SPAB~\alpha~\beta = \bigcap_{\alpha'<\alpha} \bigcup_{\beta'<\beta}
    (A \to \SPAB~\alpha~\beta') + (B \times \SPAB~\alpha'~\infty)
\]
where $\infty$ represents the closure ordinal.  The nesting is now
defined by the lexicographic recursion pattern, so we do not need to
represent it in the order of quantifiers.  Pushing them in maximally
yields an alternative definition:
\[
  \SPAB~\alpha~\beta =  
    (A \to \bigcup_{\beta'<\beta} \SPAB~\alpha~\beta') + 
    (B \times \bigcap_{\alpha'<\alpha} \SPAB~\alpha'~\infty)
\]
This variant is close to the mixed data types of Agda
\cite{danielssonAltenkirch:mpc10}, where recursive occurrences are
inductive unless marked with $\infty$:
\begin{quotecode}
\begin{alltt}
\textbf{data} SP (A B : \textbf{Set}) : \textbf{Set} \textbf{where}
  get : (A \(\to\) SP A B) \(\to\) SP A B
  put : B \(\to\) \(\infty\) (SP A B) \(\to\) SP A B 
\end{alltt}
\end{quotecode}
In Agda, one cannot specify the nesting order, it always considers the
greatest fixed-point to be on the outside
\cite{altenkirchDanielsson:par10}.

Let us program with mixed types via bounded
quantification in MiniAgda!  The type of stream processors is defined
recursively, with lexicographic termination measure
\lstinline+|i,j|+.  The bounded existential $\exists j'<j. T$ has
concrete syntax \lstinline|[j' < j] & T|, and \lstinline|Either X Y|
with constructors \lstinline|left: X ->   Either X Y| and 
\lstinline|right : Y ->   Either X Y| is the (definable) disjoint sum
type.  We directly code the ``mixed'' definition of $\SP$:
\begin{code}
\begin{lstlisting}
cofun SP : -(A : Set) +(B : Set) -(i : Size) +(j : Size) -> |i,j| -> Set
{ SP A B i j = Either     (A -> [j' < j] &  SP A B i j')
                          (B & ([i' < i] -> SP A B i' #))
} 
pattern get f       = left  f 
pattern put b sp    = right (b , sp)
\end{lstlisting}
\end{code}
We can \emph{run} a stream processor of depth $i$ and height $j$
on an $A$-stream of unbounded depth
($\infty$) to yield a $B$-stream of depth $i$ 
(this is also called stream \emph{eating}
\cite{ghaniHancockPattinson:lmcs09}).  If the stream processor is a
$\tget~f$, we feed the head of the stream to $f$, getting an new
stream processor of smaller height (index $j$), and continue running
on the stream tail.  If the stream processor is a $\tput~b~\vsp$,
we produce a $\lambda i'<i$ delayed stream whose head is $b$ and tail 
is computed by
running $\vsp$, which has smaller depth (index $i$) but unbounded
height (index $j$).
\begin{code}
\begin{lstlisting}
cofun run :   [A, B : Set] [i, j : Size] -> |i,j| -> SP A B i j -> Stream A # -> Stream B i
{ run A B i j (get f)        as = case f (head A # as) 
                                   { (j', sp) -> run A B i j' sp (tail A # as) }
; run A B i j (put b sp)     as = \ i' -> (b, run A B i' # (sp i') as) 
} 
\end{lstlisting}
\end{code}
A final note on quantifier placement:  For monotone $F$ and
$\MU^\alpha = F~(\bigcup_{\beta<\alpha} \MU^\beta)$ we have
$\MU^\alpha F = \mu^{\alpha+1} F$.  In particular $\MU^0 F = F \bot$,
thus for the list generator $F~X = \top + A \times X$ the first
approximant $\MU^0 F$ is not empty but contains exactly the empty
list.  Type $\MU^\alpha F$ contains the lists of maximal length
$\alpha$.  This encoding of data type approximants is more suitable
for size arithmetic and has been advocated by Barthe, Gr\'egoire, and
Riba \cite{bartheGregoireRiba:csl08}; in practice, it might be
superior---time will tell.


\section{Conclusions}
\label{sec:concl}

We have given a short introduction into a type system
for termination based on ordinal iteration.  Bounded size
quantification, inspired by inflationary fixed points, and recursion with
ordinal lexicographic termination measures are sufficient to encode
inductive and coinductive types and recursive and corecursive
definitions and all mixings thereof.  The full power of classical
ordinals is not needed to justify our recursion schemes:  We only need
a well-founded order $<$ that is ``long enough'' and has a successor
operation.  
I conjecture that 
set induction or constructive ordinals
(Aczel and Rathjen~\cite{aczelRathjen:constructiveSetTheory})
can play this role, leading to a constructive justification of
type-based termination. 

While our prototype MiniAgda lacks type reconstruction needed for
an enjoyable programming experience, it is evolving into a core language for
dependent type theory with termination certificates.  Our long-term goal is to
extend Agda with type-based termination in a way that most termination
certificates will be constructed automatically.  MiniAgda could serve
as an intermediate language that double-checks proofs constructed by
Agda, erases static code, and feeds the rest into a compiler
back-end.

\paradot{Acknowledgements}
I am grateful for discussions with Cody Roux which exposed a problem
with MiniAgda's pattern matching and set me on the track towards
bounded quantification as basic principle for type-based termination.
Thanks to Brigitte Pientka for many discussions on sized types and the
invitation to McGill, where some ideas of this paper prospered.
Finally, I thank the MiniAgda users, especially Nils Anders Danielsson
and David Thibodeau, who have coped with the user-unfriendliness of the
system and kept me busy fixing bugs.

\bibliographystyle{eptcsalpha}
\bibliography{auto-fics12}

\end{document}